\documentclass[showpacs,twocolumn,pra]{revtex4}
\usepackage{graphicx,psfrag}
\begin{document}

\title{Motion-light parametric amplifier and entanglement distributor}
\author{A. Peng}
\author{A.~S. Parkins}
\email[Corresponding author. Email address: ]{s.parkins@auckland.ac.nz}
\affiliation{Department of Physics, University of Auckland,
Private Bag 92019, Auckland, New Zealand.}
\date{\today}

\begin{abstract}
We propose a scheme for entangling the motional mode of a trapped atom
with a propagating light field via a cavity-mediated parametric
interaction.
We then show that if this light field is subsequently coupled to a second
distant atom via a cavity-mediated linear-mixing interaction,
it is possible to transfer the entanglement from the light beam to the
motional mode of the second atom to create an EPR-type entangled state of
the positions and momenta of two distantly-separated atoms.
\end{abstract}

\maketitle

\section{Introduction}
The generation, distribution, and application of continuous variable
quantum entanglement are topics of considerable interest at present,
spurred on in large part by the burgeoning fields of quantum communication
and quantum computation.
In this context, a variety of protocols for continuous quantum variables
have been proposed and in some cases already demonstrated, including
quantum teleportation \cite{Vaidman94,Braunstein98,Furusawa98},
quantum cryptography \cite{Ralph00,Hillery00,Pereira00,Reid00},
quantum dense coding \cite{Ban99,Braunstein00,Li02}, and
quantum computation \cite{Lloyd99}.

These protocols have in large part focussed on implementations
involving nonlinear optics and propagating
light fields, which has obvious advantages in terms of
long-distance communication and existing quantum-optical technology.
However, for purposes of storage (i.e., memory) and local manipulation,
a number of alternative physical systems are being actively investigated
and appear very promising; notably,
collective atomic spin systems
\cite{Lukin00,Mair02,Turukhin02,Zibrov02,Kozhekin00,Duan00,Julsgaard01}
and quantized vibrational states of trapped atoms
\cite{Parkins99,Parkins00a,Parkins00b,Parkins01,Mancini01}.
Of particular interest is the capability of establishing long-lived
entanglement of the Einstein-Podolsky-Rosen (EPR) type \cite{Einstein35}
between actual or effective position and momentum variables of two
separated atomic systems. This capability has in fact been demonstrated
recently in an experiment involving collective spins of a pair of atomic
ensembles and nonlocal Bell measurements using off-resonant light pulses
\cite{Julsgaard01}.

A scheme for preparing an EPR-type state of the actual positions and
momenta of a pair of atoms has been put forward in \cite{Parkins00a},
making use of interactions in cavity quantum electrodynamics (cavity
QED) to facilitate the transfer of quantum correlations from light
fields to motional modes of tightly trapped atoms. The source of the
quantum-correlated light fields was taken to be an optical nondegenerate
parametric amplifier (see, e.g., \cite{Ou92a,Ou92b}).

Here, we present an alternative approach to preparing such a motional
state which, in contrast to the scheme of
\cite{Parkins00a}, does not require a separate source of
quantum-correlated light beams. Furthermore, unlike a number of other
proposals, it does not require entangling measurements to be made, or a
carefully timed sequence of suitably shaped light pulses.
Through an atom-cavity coupling similar to, but modified from that of
\cite{Parkins00a}, an effective parametric interaction between cavity
and motional modes generates continuous variable entanglement between
the motion of the trapped atom and the light field exiting the cavity.
The entanglement ``carried'' by this light field can subsequently be
distributed to a distant location and thence to another atom
(or atoms).

\section{The Model}

The essential details of the scheme to be described in this work are
illustrated in Fig.~1. A pair of atoms are harmonically confined inside
separate optical cavities, with the light exiting one of the cavities
coupled into the second cavity (but not vice-versa). Auxiliary lasers,
incident through the sides of the cavities, combine with the cavity
fields to drive Raman transitions between neighbouring vibrational
levels of the motion of each atom. Note that only a single internal
atomic state (i.e., a stable ground electronic state) is assumed to
be involved, as will be discussed below.

\begin{figure}[h]
\begin{psfrags}
    \psfrag{mu 1x}{$\nu_{1x}$}
    \psfrag{mu 2x}{$\nu_{2x}$}
    \psfrag{n 1x}{$n_{1x}$}
    \psfrag{n 2x}{$n_{2x}$}
\includegraphics[width=8cm]{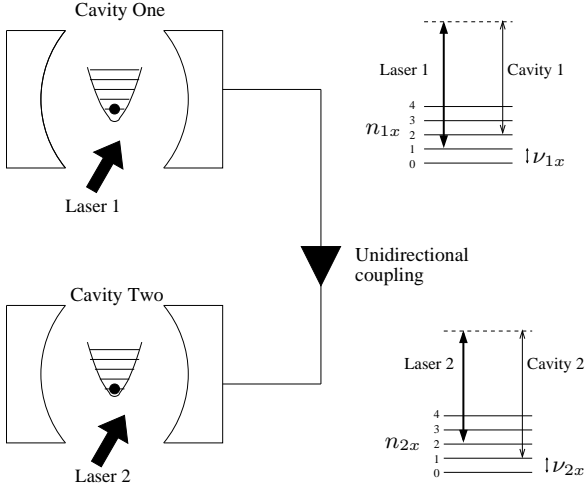}
\caption{Schematic diagram of the system. Atoms inside the cavities are
tightly confined in harmonic traps. The cavity and laser frequencies are
chosen to drive Raman transitions between neighbouring vibrational levels
of the motion.}
\end{psfrags}
\end{figure}

\subsection{Motion-Light Coupling}

The laser-atom-cavity interactions responsible for the coupling between
motional and light modes have been discussed in detail previously
(see, e.g., \cite{Parkins99,Parkins01}), but for completeness we include
a brief description.
Considering just a single-atom configuration, our system is
modeled, in a frame rotating at the laser frequency $\omega_{\rm L}$,
by the Hamiltonian
\begin{eqnarray}
H_{0}&=&\sum_{j=x,y,z}\hbar\nu_{j}(b_{j}^{\dagger}b_{j}
+\frac{1}{2})\nonumber
\\
&+&\hbar\delta a^{\dagger}a
+\hbar\Delta\sigma_{+}\sigma_{-} \nonumber
\\
&+& \hbar[ {\cal E}_{\rm L}(y,z,t)\sigma_{+}+{\cal E}_{\rm L}^\ast (y,z,t)
\sigma_{-}] \nonumber
\\
&+& \hbar g_{0} \sin (kx)(a^{\dagger}\sigma_{-}
+a \sigma_{+}) .
\label{eq:H0}
\end{eqnarray}
The first line describes the quantized harmonic motion of the trapped atom,
with $b_j$ the annihilation operator and $\nu_j$ the vibrational frequency
for motion along the $j$-axis.
The quantized cavity mode, with annihilation operator $a$ and frequency
$\omega_{\rm cav}$, is detuned from the laser frequency by
$\delta =\omega_{\rm cav}-\omega_{\rm L}$.
The ground and excited electronic states of the atom that are coupled
by the light fields, $|g\rangle$ and $|e\rangle$, are separated in energy
by $\hbar\omega_0$, and the detuning of the atomic transition from the laser
frequency is given by $\Delta =\omega_0-\omega_{\rm L}$;
$\sigma_+=|e\rangle\langle g|$ and $\sigma_-=|g\rangle\langle e|$ are the
atomic raising and lowering operators, respectively.

The laser field is treated as a classical field of (complex) amplitude
${\cal E}_{\rm L}(y,z,t)$ and is assumed to propagate in the $yz$-plane.
The cavity mode is aligned along the $x$-axis and its coupling to the
atomic transition is described by the last line in (\ref{eq:H0}), where
$g_{0}$ is the single-photon coupling strength and $k$ is the wavenumber
of the cavity field. The choice of a sine function, with
$x=(\hbar /2m\nu_x)^{1/2}(b_x+b_x^\dagger )$ the position operator of the
atom, denotes that the trap is assumed to be centered at a {\it node} of
the cavity standing-wave field.

By taking the laser-atom detuning $\Delta$ to be large
(i.e., $\Delta\gg\{\nu_x,\delta ,|{\cal E}_{\rm L}|,g_0,\gamma\}$,
where $\gamma$ is the spontaneous emission linewidth of the state
$|e\rangle$), population of the excited internal state $|e\rangle$,
and hence spontaneous emission, can be made negligible.
From the Heisenberg equation of motion for the atomic lowering operator,
\begin{equation}
\dot{\sigma}_{-} = -i\Delta\sigma_{-}
+ i{\cal E}_{\rm L}(y,z,t)\sigma_{z}
+ ig_{0}\sin (kx)a\sigma_{z} ,
\end{equation}
one can then take
\begin{equation} \label{eq:sigmam}
\sigma_{-} \simeq \frac{-{\cal E}_{\rm L}(y,z,t)}{\Delta}
-\frac{g_{0}\sin (kx)a}{\Delta} ,
\end{equation}
noting that $\dot{\sigma}_{z}\simeq 0$ and
$\sigma_z|g\rangle =-|g\rangle$ (i.e., setting $\sigma_z\simeq -1$ in
the equation of motion for $\sigma_-$).

Now, if the size of the harmonic trap is small compared to the optical
wavelength (Lamb-Dicke regime), then we can make the approximation
\begin{equation}
\sin (kx) \simeq \eta_{x}(b_{x}+b_{x}^{\dagger}) ,
\end{equation}
where $\eta_x\ll 1$ is the Lamb-Dicke parameter. In this case,
one can also assume a configuration such that the position dependence
of the laser amplitude over the extent of the trap can be ignored, i.e.,
\begin{equation}
{\cal E}_{\rm L}(y,z,t) \simeq {\cal E}_{\rm L}(t) e^{-i\phi_{\rm L}} ,
\;\;\;\;  {\cal E}_{\rm L}(t) \in \Re \, ,
\end{equation}
and hence only motion along the $x$-axis is of relevance.

Finally, for the regimes we are most interested in, the cavity mode is
only ever weakly excited. This fact, combined with the smallness of the
Lamb-Dicke parameter, enables us to neglect the second term in
(\ref{eq:sigmam}) in comparison to the first term, reducing our
approximate solution for $\sigma_-$ to the simple form
\begin{equation}
\sigma_- \simeq - \frac{{\cal E}_{\rm L}(t) e^{-i\phi_{\rm L}}}{\Delta} .
\end{equation}
Using this form, and shifting the zero of energy to remove scalar
terms, the Hamiltonian describing the cavity mode and motion along the
$x$-axis reduces to \cite{Zeng94}
\begin{eqnarray} \label{eq:H0approx}
H_{0} &\simeq & \hbar
\nu_{x}b_{x}^{\dagger}b_{x} +
\hbar\delta a^{\dagger}a \nonumber
\\
& &- \frac{\hbar\eta_{x}g_{0}{\cal E}_{\rm L}(t)}{\Delta}
(b_{x}+b_{x}^{\dagger}) (a^{\dagger}
e^{-i\phi_{\rm L}}+ae^{i\phi_{\rm L}}) .
\end{eqnarray}
Under suitable conditions and
with appropriate choices of the cavity-laser detuning,
$\delta =\omega_{\rm cav}-\omega_{\rm L}$, we are able to
choose between a parametric or linear mixing interaction
between the cavity and motional modes, as we now show.

\subsection{Cavity 1: Parametric Amplification}

For convenience, we now drop the subscripts $x$ and ${\rm L}$
and use the subscript $1$(2) to denote system 1(2).
For cavity 1 we choose $\delta_1 =-\nu_1$. Moving to an
interaction picture to remove the systematic motion associated
with the first two terms of (\ref{eq:H0approx}), we obtain
\begin{eqnarray} \label{eq:pa1}
H_{1} &=& \Omega_1(t)
\left( a_{1}b_{1}e^{i\phi_1} + a_{1}^{\dagger}b_{1}^{\dagger}
e^{-i\phi_1} \right. \nonumber
\\
&& \left. \;\;\;\;\;\; + \,
a_{1}^\dagger b_{1}e^{-2i\nu_1t-i\phi_1} + a_{1}b_{1}^{\dagger}
e^{2i\nu_1t+i\phi_1} \right) ,
\end{eqnarray}
where
\begin{equation}
\Omega_1(t) = -\frac{\hbar\eta_1g_1{\cal E}_1(t)}{\Delta_1} .
\end{equation}
If we now assume that the trap frequency $\nu_1$ is large, such that
$\nu_1\gg |\Omega_1(t)|$, and also $\nu_1\gg\kappa_1$, where $\kappa_1$
is the amplitude decay rate of the cavity field, then the rapidly
oscillating terms in (\ref{eq:pa1}) can be neglected in a rotating-wave
approximation to give
\begin{equation} \label{eq:pa2}
H_{1} = \Omega_1(t)
\left( a_{1}b_{1}e^{i\phi_1} + a_{1}^{\dagger}b_{1}^{\dagger}
e^{-i\phi_1} \right) .
\end{equation}
This is the Hamiltonian for a nondegenerate parametric amplification
process and via this process continuous variable entanglement
can be generated between the motional and cavity modes in system 1.

\subsection{Cavity 2: Linear Mixing}

For cavity 2 we choose $\delta_2=\nu_2$. Moving to the appropriate
interaction picture and employing the rotating-wave approximation once
again under the condition that
$\nu_2\gg |\Omega_2(t)|$ and $\nu_2\gg\kappa_2$, we derive
for system 2 the effective Hamiltonian
\begin{equation} \label{eq:lm}
H_{2} = \Omega_2(t)
\left( a_{2}^\dagger b_{2}e^{-i\phi_2} + b_{2}^{\dagger}a_{2}
e^{i\phi_2} \right) ,
\end{equation}
where
\begin{equation}
\Omega_2(t) = -\frac{\hbar\eta_2g_2{\cal E}_2(t)}{\Delta_2} .
\end{equation}
This describes a linear mixing interaction which, as we shall see,
can lead to an exchange of properties between the cavity and motional
modes.

\subsection{Cascaded System}

Having specified the effective interactions occurring in each
atom-cavity system, it should now be quite clear what our aim is.
Via the parametric interaction, light in cavity mode 1 is entangled
with motional mode 1. This light ultimately exits cavity 1 due
to cavity decay at the rate $\kappa_1$ and it may then be coupled
into cavity 2, where, through the linear mixing process, entanglement
of the light field with motional mode 1 may be transferred to
motional mode 2, thereby entangling the two motional modes.

To examine this transfer process in detail, we turn naturally to a
cascaded systems model
\cite{Gardiner93,Carmichael93,Kochan94,Gardiner94,Gardiner00},
which enables us to describe a unidirectional driving of cavity
2 with the output light from cavity 1.
If we assume that the dominant input and output channel to and
from each cavity is through just one of the two mirrors forming
each cavity (as depicted in Fig.~1), then the master equation
for our cascaded system is
\begin{eqnarray} \label{eq:mecas}
\dot{\rho}=&-&\frac{i}{\hbar}[H,\rho] \nonumber
\\
&+& \kappa_{1}(2a_{1}\rho a_{1}^{\dagger} - a_{1}^{\dagger}a_{1}\rho
- \rho a_{1}^{\dagger}a_{1}) \nonumber
\\
&+&\kappa_{2}(2a_{2}\rho a_{2}^{\dagger} - a_{2}^{\dagger}a_{2}\rho
- \rho a_{2}^{\dagger}a_{2}) \nonumber
\\
&-&2\sqrt{\epsilon \kappa_{1}\kappa_{2}}([a_{2}^{\dagger},a_{1}\rho]
+ [\rho a_{1}^{\dagger},a_{2}]) ,
\end{eqnarray}
where
\begin{equation}
H = H_1 + H_2 ,
\end{equation}
and the parameter $\epsilon$, satisfying $0\leq\epsilon\leq 1$, accounts
for losses in transmission and for coupling inefficiency.
Ideal transmission and coupling corresponds to $\epsilon =1$.
Note that implicit in the form (\ref{eq:mecas}) is the assumption that
the two cavity frequencies are equal, and hence that the difference in
laser frequencies is set to
\begin{equation}
\omega_{\rm L1} - \omega_{\rm L2} = \nu_1 + \nu_2 .
\end{equation}

\section{Adiabatic Approximation}

In the (overdamped) regime where both $\kappa_{1}$ and $\kappa_{2}$
are much larger than the coupling rates $\Omega_{1}$ and
$\Omega_{2}$, our model can be simplified further by adiabatically
eliminating the cavity modes from the dynamics.
In particular, the master equation (4) can be written in the form
\begin{equation}
\dot{\rho} = (\mathcal{L}_{0} + \mathcal{L}_{\rm c}) \rho \, ,
\end{equation}
where
\begin{eqnarray}
\mathcal{L}_{0}\rho &=& \frac{-i}{\hbar}[H,\rho] ,
\\
\mathcal{L}_{\rm c}\rho &=& \kappa_{1}(2a_{1}\rho a_{1}^{\dagger}
- a_{1}^{\dagger}a_{1}\rho - \rho a_{1}^{\dagger}a_{1}) \nonumber
\\
&+&\kappa_{2}(2a_{2}\rho a_{2}^{\dagger} - a_{2}^{\dagger}a_{2}\rho
- \rho a_{2}^{\dagger}a_{2}) \nonumber
\\
&-&2\sqrt{\epsilon \kappa_{1}\kappa_{2}}([a_{2}^{\dagger},a_{1}\rho]
+ [\rho a_{1}^{\dagger},a_{2}]) .
\end{eqnarray}
In the adiabatic limit, a master equation for the reduced density operator,
$\rho_b$, of the motional modes alone can be derived as
(see, e.g., \cite{Gardiner00})
\begin{equation} \label{eq:drhobdt}
\dot{\rho_{b}} = {\rm Tr}_{\rm c} \left\{ \mathcal{L}_{0}\int_{0}^{\infty} d\tau\,
e^{-\mathcal{L}_{\rm c}\tau} \mathcal{L}_{0} \rho_{\rm c}^{\rm s} \right\}
\rho_{b} \, ,
\end{equation}
where the trace is taken over the cavity modes and
$\rho_{\rm c}^{\rm s}$ is the steady state density operator of
the cavity modes satisfying ${\cal L}_{\rm c}\rho_{\rm c}^{\rm s}=0$.

Evaluation of (\ref{eq:drhobdt}) requires calculation of steady-state two-time
correlation functions of the cavity operators.
From the master equation $\dot{\rho}_{\rm c}=\mathcal{L}_{\rm c}\rho_{\rm c}$
one obtains the following solutions for the mean cavity amplitudes
(for $t\geq 0$)
\begin{eqnarray}
\langle a_{1}(t) \rangle_{\rm c} &=& e^{-\kappa_{1}t}
\langle a_{1}(0)\rangle_{\rm c} ,
\\
\langle a_{2}(t) \rangle_{\rm c} &=& e^{-\kappa_{1}t} \langle
a_{2}(0)\rangle_{\rm c} \nonumber
\\
&& \; +
\frac{2\sqrt{\kappa_{1}\kappa_{2}\epsilon}}{\kappa_{2}-
\kappa_{1}}(e^{-\kappa_{2}t}-e^{-\kappa_{1}t})\langle
a_{1}(0)\rangle_{\rm c} .
\end{eqnarray}
With $\langle a_{1}a_{1}^{\dagger}\rangle_{\rm c}^{\rm s} = \langle
a_{2}a_{2}^{\dagger}\rangle_{\rm c}^{\rm s}=1$ being the only nonzero
steady-state equal-time correlations, the quantum regression theorem gives
\begin{eqnarray}
\langle a_{1}(\tau)a_{1}^{\dagger}(0) \rangle_{\rm c}^{\rm s} &=&
\langle a_{1}(0)a_{1}^{\dagger}(\tau) \rangle_{\rm c}^{\rm s} = e^{-\kappa_{1}\tau} ,
\\
\langle a_{2}(\tau)a_{2}^{\dagger}(0) \rangle_{\rm c}^{\rm s} &=&
\langle a_{2}(0)a_{2}^{\dagger}(\tau) \rangle_{\rm c}^{\rm s} = e^{-\kappa_{2} \tau} ,
\\
\langle a_{2}(\tau)a_{1}^{\dagger}(0) \rangle_{\rm c}^{\rm s} &=& \langle
a_{1}(0)a_{2}^{\dagger}(\tau)\rangle_{\rm c}^{\rm s} \nonumber
\\
&=&
\frac{2\sqrt{\kappa_{1}\kappa_{2}\epsilon}}{\kappa_{2}-
\kappa_{1}}(e^{-\kappa_{2}\tau}-e^{-\kappa_{1}\tau}) ,
\end{eqnarray}
as the only non-zero two-time correlation functions (resulting from the
master equation $\dot{\rho}_{\rm c}=\mathcal{L}_{\rm c}\rho_{\rm c}$).

Using these correlation functions in (\ref{eq:drhobdt}),
the master equation for the reduced density operator of the {\em
motional modes} is
\begin{eqnarray} \label{eq:merhob}
\dot{\rho_{b}}
&=&\Gamma_{1}(2b_{1}^{\dagger}\rho_{b} b_{1} - b_{1}b_{1}^{\dagger}\rho_{b}
- \rho_{b} b_{1}b_{1}^{\dagger}) \nonumber
\\
&+&\Gamma_{2}(2b_{2}\rho_{b} b_{2}^{\dagger} - b_{2}^{\dagger}b_{2}\rho_{b}
- \rho_{b} b_{2}^{\dagger}b_{2}) \nonumber
\\
&+&2\sqrt{\epsilon \Gamma_{1}\Gamma_{2}} \left(
[b_{2}^{\dagger},b_{1}^{\dagger}\rho_{b}] e^{-i(\phi_1-\phi_2)} \right.
\nonumber
\\
&& \;\;\;\;\;\;\;\;\;\;\;\;\;\;\;\;\; + \left.
[\rho_{b}b_{1},b_{2}] e^{i(\phi_1-\phi_2)} \right)  ,
\end{eqnarray}
where
\begin{equation}
\Gamma_{1} = \frac{\Omega_{1}^{2}}{\kappa_{1}} \;\; \textrm{and} \;\;
\Gamma_{2} = \frac{\Omega_{2}^{2}}{\kappa_{2}}
\end{equation}
are the effective growth and decay rates, respectively, for motional modes
1 and 2.
In this reduced model a direct coupling between these modes now appears
in the form of the last term in (\ref{eq:merhob}).

\section{Motional State Entanglement}

\subsection{Motional Mode Correlations}

From the master equation (\ref{eq:merhob}), a closed set of (inhomogeneous)
differential equations can be obtained
for the correlation functions $\langle b_{1}^{\dagger}b_{1}\rangle$,
$\langle b_{2}^{\dagger}b_{2}\rangle$ and $\langle b_{1}b_{2}\rangle$.
These are (setting $\phi_1=\phi_2$ for simplicity)
\begin{eqnarray}
\dot{\langle b_{1}b_{2}\rangle}&=&(\Gamma_{1} - \Gamma_{2})\langle
b_{1}b_{2}\rangle \nonumber
\\
&& \;\; + 2\sqrt{\epsilon \Gamma_{1}\Gamma_{2}}\,
(\langle b_{1}^{\dagger}b_{1}\rangle + 1) , \\
\dot{\langle b_{1}^{\dagger}b_{1}\rangle}&=& 2\Gamma_{1}(1+\langle
b_{1}^{\dagger}b_{1}\rangle) , \\
\dot{\langle b_{2}^{\dagger}b_{2}\rangle}&=& -2\Gamma_{2}\langle
b_{2}^{\dagger}b_{2}\rangle \nonumber
\\
&& \;\; + 2\sqrt{\epsilon
\Gamma_{1}\Gamma_{2}}(\langle b_{1}^{\dagger}b_{2}^{\dagger}\rangle
+ \langle b_{1}b_{2}\rangle) ,
\end{eqnarray}
with solutions
\begin{eqnarray}
\langle b_{1}^{\dagger}(t)b_{1}(t)\rangle &=& e^{2\Gamma_{1}t} - 1 ,
\\
\langle b_{1}(t)b_{2}(t)\rangle &=& \frac{2\sqrt{\epsilon
\Gamma_{1}\Gamma_{2}}}{\Gamma_{1}+\Gamma_{2}}e^{\Gamma_{1}t}
(e^{\Gamma_{1}t} - e^{-\Gamma_{2}t}) , \\
\langle b_{2}^{\dagger}(t)b_{2}(t)\rangle &=& \frac{4\epsilon
\Gamma_{1}\Gamma_{2}}{(\Gamma_{1}+\Gamma_{2})^{2}}(e^{\Gamma_{1}t}
- e^{-\Gamma_{2}t})^{2} ,
\end{eqnarray}
where both oscillators are assumed to have initially been in their
ground states.
Note that, under these conditions,
$\langle b_{1}(t)b_{1}(t)\rangle =\langle b_{2}(t)b_{2}(t)\rangle =
\langle b_{1}^{\dagger}(t)b_{2}(t)\rangle = 0$ for all $t$.

The growing value of $\langle b_{1}(t)b_{2}(t)\rangle$ demonstrates
that our cascaded system enables correlations to develop
between the two motional modes. The nature of the correlations is,
not surprisingly, reminiscent of a two-mode squeezed state and
leads us to examine correlations between the positions and momenta
of the trapped atoms.

Before proceeding, however, we note that the exponential growth
exhibited by the mean excitation numbers must eventually become
problematical for our model, since the Lamb-Dicke assumption
breaks down once the excitation numbers become large
enough that the physical extent of the atomic wavepacket is no longer
much smaller than the wavelength of the light. We shall return to
this point when we consider practical aspects of the scheme.

\subsection{Position and Momentum Variances}

Defining position and momentum operators for the motional
modes as
\begin{eqnarray}
X_{j} = b_{j}+b_{j}^{\dagger} , \;\;\;
P_{j} = -i(b_{j}-b_{j}^{\dagger}) , \;\;\;\; (j=1,2) ,
\end{eqnarray}
variances in the sum and difference operators are calculated to be
\begin{eqnarray}
&& \langle (X_{1} \pm X_{2})^{2} \rangle = \langle (P_{1}
\mp P_{2})^{2}\rangle \nonumber \\
&& \;\;\;\;\;\; = 2\left[ e^{\Gamma_{1}t} \pm \frac{2\sqrt{\epsilon
\Gamma_{1}\Gamma_{2}}}{\Gamma_{1}+\Gamma_{2}}(e^{\Gamma_{1}t}-
e^{-\Gamma_{2}t})\right]^{2} ,
\end{eqnarray}
where the ``vacuum'' level (i.e., where the atoms are both in
their ground motional states) is $2$.
We note first that, since $(e^{\Gamma_{1}t}-e^{-\Gamma_{2}t})\geq 0$,
$\langle (X_1+X_2)^2\rangle =\langle (P_1-P_2)^{2}\rangle\geq 2$
at all times, i.e., these variances are only ever increased compared
to the vacuum level.

However, the variances $\langle (X_{1}-X_{2})^{2} \rangle$ and
$\langle (P_{1}+P_{2})^{2} \rangle$ {\em can} be reduced below the value
of 2, as illustrated in Fig.~2, where
$\langle (X_{1}-X_{2})^{2} \rangle$ is plotted as a function of time
for various values of the ratio $\Gamma_2/\Gamma_1$, and with
$\epsilon =1$.
The case $\Gamma_2/\Gamma_1=1$, where the
effective damping rates of the two motional modes are equal,
is particularly interesting and leads to the simplified result
\begin{eqnarray}
\langle(X_{1}-X_{2})^{2}\rangle &=& \langle(P_{1}+P_{2})^{2}\rangle
\nonumber
\\
&=& 2e^{-2\Gamma_{1}t}\rightarrow 0 \;\;\;  \textrm{as} \;\;\;
t\rightarrow \infty  .
\end{eqnarray}
So, using this scheme it is, in principle, possible to produce a
state corresponding to
perfectly correlated positions and perfectly anti-correlated momenta
of the two atoms, i.e., to produce an EPR state.
Once again, in comparing this approach to that
of \cite{Parkins00a}, it should be emphasized that the present scheme
{\em does not require a separate source of entangled light beams}.

\begin{figure}[h]
\begin{center}
\includegraphics[scale=0.42]{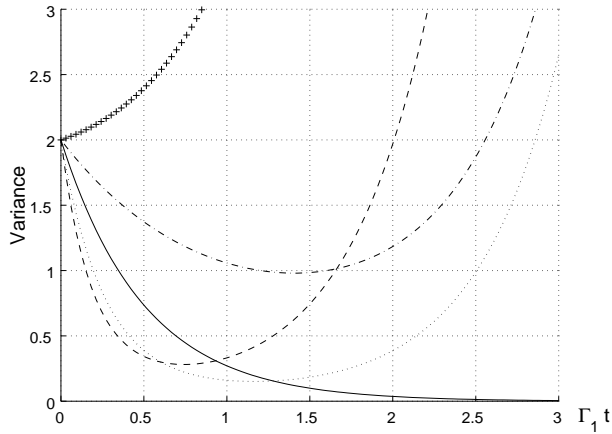}
\caption{The variance $\langle (X_1-X_2)^2\rangle$ as a function of
time for $\Gamma_2/\Gamma_1=0.2$ (pluses),
$\Gamma_2/\Gamma_1=0.5$ (dot-dashed line), $\Gamma_2/\Gamma_1=1$
(solid line), $\Gamma_2/\Gamma_1=2$ (dotted line), and
$\Gamma_2/\Gamma_1=3$ (dashed line).
The coupling is taken to be ideal, i.e., $\epsilon =1$.}
\end{center}
\end{figure}

If $\Gamma_2/\Gamma_1\neq 1$ and/or $\epsilon <1$, then the variance
$\langle (X_1-X_2)^2\rangle$ attains a finite minimum value
at a particular time, after which it increases indefinitely.
Defining $\lambda =\Gamma_2/\Gamma_1$, this
minimum value is calculated to be
\begin{widetext}
\begin{equation}
\langle (X_{1}-X_{2})^{2}\rangle_{\rm min} = \frac{2}{(1+\lambda)^2}
\left[  \frac{(2\sqrt{\epsilon}\lambda^{\frac{3}{2}})^{\frac{1}{1
+\lambda}}}{(1+\lambda-2\sqrt{\lambda\epsilon})^{\frac{1}{1
+\lambda}-1}}+2\sqrt{\lambda\epsilon}\left(\frac{2
\sqrt{\epsilon}\lambda^{\frac{3}{2}}}{1+\lambda-2
\sqrt{\lambda\epsilon}}\right)^{\frac{-\lambda}{1
+\lambda}}\right]^{2} ,
\end{equation}
\end{widetext}
and the time at which the minimum value occurs is given by
\begin{equation}
\Gamma_1t_{\rm min}(\lambda )=\frac{1}{1+\lambda} \ln
\left[\frac{2\lambda\sqrt{\lambda\epsilon}}{1
+\lambda -2\sqrt{\lambda\epsilon}}\right] .
\end{equation}
Plots of $\langle (X_{1}-X_{2})^{2}\rangle_{\rm min}$
and $\Gamma_1t_{\rm min}$ versus $\lambda$
are shown in Figs.~3 and 4 for several values of $\epsilon$.
Note that $t_{\rm min}>0$ only where $\lambda >1/(4\epsilon )$.
If $\lambda <1/(4\epsilon )$ then there is no reduction in the
variance below the vacuum level.

\begin{figure}[h]
\begin{center}
\includegraphics[scale=0.42]{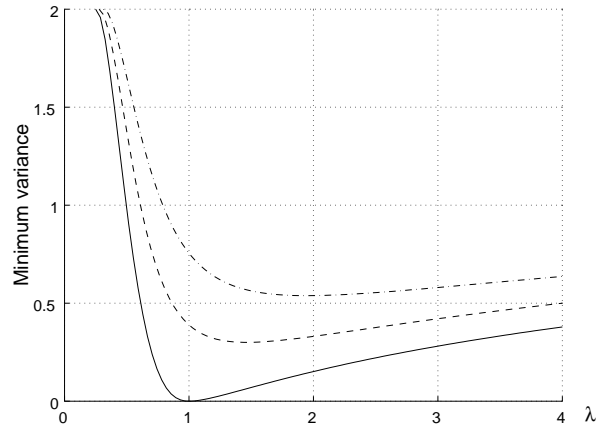}
\caption{Minimum variance
$\langle (X_{1}-X_{2})^{2} \rangle_{\rm min}$ as a function
of $\lambda =\Gamma_2/\Gamma_1$
for $\epsilon=1$ (solid line), $\epsilon=0.9$ (dashed line)
(minimum value of $0.30$ at $\lambda=1.46$), and $\epsilon=0.8$
(dot-dashed line) (minimum value of $0.54$ at $\lambda=1.95$).}
\end{center}
\end{figure}

\begin{figure}[h]
\begin{center}
\includegraphics[scale=0.42]{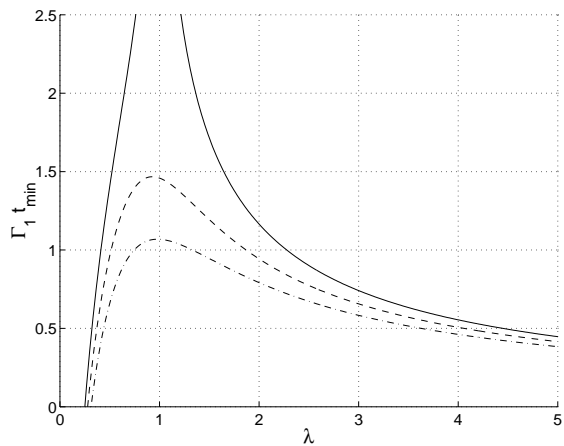}
\caption{Time at which minimum variance occurs,
$\Gamma_1t_{\rm min}$, as a function of
$\lambda =\Gamma_2/\Gamma_1$ for $\epsilon=1$ (solid line),
$\epsilon=0.9$ (dashed line),
and $\epsilon=0.8$ (dot-dashed line).}
\end{center}
\end{figure}

From Figs.~3 and 4 (in particular, from their asymmetry about the
point $\lambda =1$),
it is clear that the scheme generally performs
best in the regime where $\Gamma_2\geq\Gamma_1$ (i.e., $\lambda\geq 1$).
In particular, significant reductions in the variances below the
vacuum level occur over a broad region of $\Gamma_2$ values,
provided $\Gamma_2\geq\Gamma_1$.

With decreasing values of the parameter $\epsilon$ (i.e., with decreasing
coupling efficiency or increasing transmission losses), the minimum
attainable variance increases and occurs for larger values of $\lambda$
and at somewhat earlier times. It follows that, for a realistic system
with $\epsilon <1$, it would be advantageous to work in a
regime with $\Gamma_2>\Gamma_1$.
This is further illustrated in Fig.~5, where the variance is again
plotted as a function of time, but now for several combinations of
$\lambda$ and $\epsilon$.
It is worth noting that a significant reduction in the variance
($\sim 73\%$) is
still possible with $\epsilon =0.8$, i.e.,
$\langle (X_1-X_2)^2\rangle_{\rm min}\simeq 0.54$ at
$\Gamma_1t_{\rm min}\simeq 0.8$ for $\lambda =2$.

It should also be noted that, for the examples chosen in the
regime where $\epsilon <1$ and $\lambda\geq 1$, the minimum variance
generally occurs at times $\Gamma_1t$ of the order of 1. The mean
excitation number for motional mode 1,
$\langle b_1^\dagger b_1\rangle$, is then of the order of
$e^2-1\simeq 6$, with a slightly smaller value for
$\langle b_2^\dagger b_2\rangle$.

\begin{figure}[h]
\begin{center}
\includegraphics[scale=0.42]{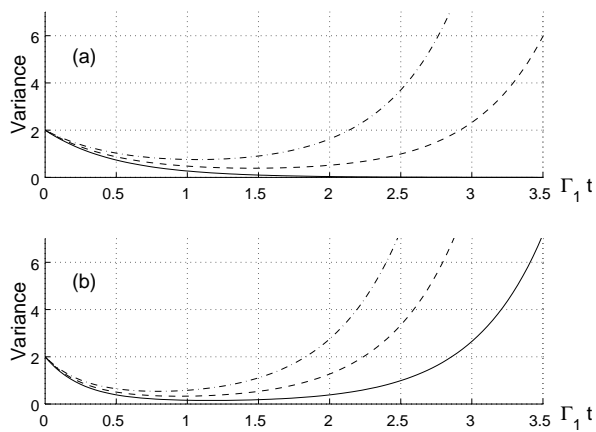}
\caption{Variance $\langle (X_{1}-X_{2})^{2} \rangle$ as a
function of time for (a) $\lambda = 1$ and
(b) $\lambda = 2$, with $\epsilon = 1$ (solid line),
$\epsilon = 0.9$ (dashed line), and $\epsilon = 0.8$
(dot-dashed line).}
\end{center}
\end{figure}

\section{Comparison With Numerical Calculations}

Starting from the master equation (\ref{eq:mecas}), which includes
the cavity dynamics, a closed set of differential equations can be
obtained for various correlation functions of the system.
We have integrated these equations numerically
(using a 4th-order Runge-Kutta method) and computed the variance
$\langle (X_{1}-X_{2})^{2} \rangle$ for comparison with the
results of the adiabatic approximation.
This comparison is presented in Figs.~6 and 7 for several example
sets of parameters.

On the timescales shown, the adiabatic approximation is seen to
work well provided the coupling strengths $\Omega_j$ are, roughly,
at least five to ten times smaller than the cavity decay rates
$\kappa_j$. For larger values of $\Omega_j$ the maximum degree
of reduction in the variance is significantly reduced and the
dynamics obviously become somewhat more complicated, e.g.,
oscillatory behaviour starts to feature and excitation of the
cavity modes increases.

\begin{figure}[h]
\begin{center}
\includegraphics[scale=0.42]{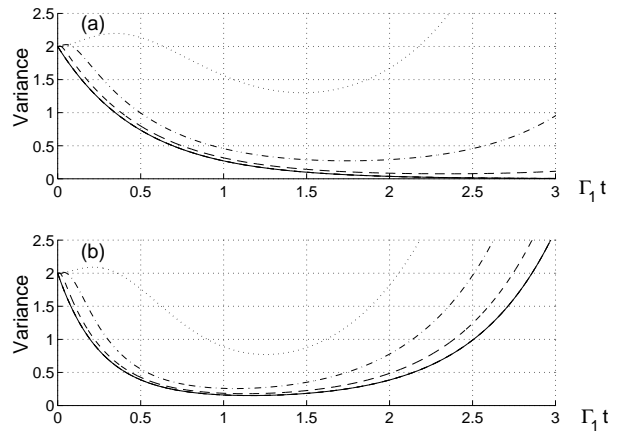}
\caption{Comparison of numerical and adiabatic approximation
results for $\langle (X_1-X_2)^2\rangle$ versus time
with $\kappa_{1}=\kappa_{2}=1$. The adiabatic approximation
is shown in each plot as a solid line.
(a) $\lambda =1$, with
$\Omega_{1}=\Omega_{2}=0.1$ (dashed line),
$\Omega_{1}=\Omega_{2}=0.2$ (dot-dashed line),
$\Omega_{1}=\Omega_{2}=0.5$ (dotted line).
(b) $\lambda =2$, with
$\Omega_{1}=0.1$, \mbox{$\Omega_{2}=\sqrt{2}\times 0.1$} (dashed line),
$\Omega_{1}=0.2$, \mbox{$\Omega_{2}=\sqrt{2}\times 0.2$} (dot-dashed line),
$\Omega_{1}=0.5$, \mbox{$\Omega_{2}=\sqrt{2}\times 0.5$} (dotted line).}
\end{center}
\end{figure}

\begin{figure}[h]
\begin{center}
\includegraphics[scale=0.42]{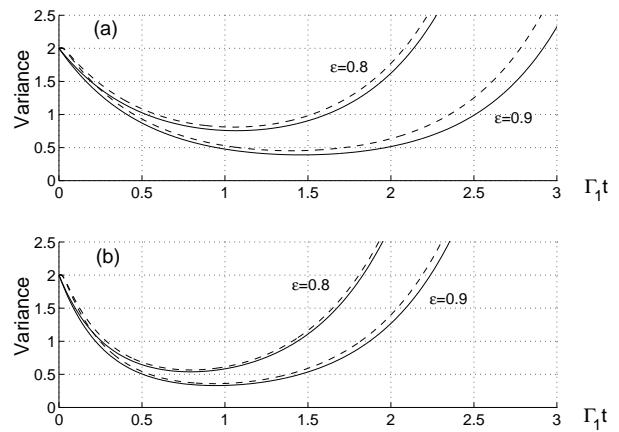}
\caption{Comparison of numerical and adiabatic approximation results
for $\langle (X_1-X_2)^2\rangle$ versus time with $\kappa_{1}=\kappa_{2}=1$
and nonideal coupling between the cavities (i.e., $\epsilon <1$).
The adiabatic approximation
is shown in each plot as a solid line.
(a) $\lambda =1$, with
$\Omega_{1}=\Omega_{2}=0.1$ (dashed lines).
(b) $\lambda =2$, with
$\Omega_{1}=0.1$, \mbox{$\Omega_{2}=\sqrt{2}\times 0.1$} (dashed lines).}
\end{center}
\end{figure}

\section{Practical Considerations}

We now consider in slightly more detail the conditions under which the
most significant assumptions required by our model should be satisfied.
Firstly though, as a more general comment, we note that exciting
progress has been made recently in experimental
cavity QED with single trapped atoms or ions
\cite{Ye99,Hoo00,Pin00,Gut01,Mun02}. In fact, various ``ingredients''
of the scheme presented in this work have already been demonstrated.

\subsection{Trap Frequency}

The neglect of terms in the effective motion-cavity mode interaction
Hamiltonians which vary like $e^{\pm 2i\nu_jt}$ requires that the trap
frequencies be large in comparison with the cavity decay rates $\kappa_j$
and the effective coupling parameters $\Omega_j$.
To quantify this a little more precisely, previous numerical studies have
shown that this rotating-wave approximation is very good provided the trap
frequencies are at least an order of magnitude larger than $\kappa_j$ and
$\Omega_j$ \cite{Parkins99,Parkins01}. We note that
an experimental situation with $\nu_j\gg\kappa$ has been realized
recently with single calcium ions trapped inside a high-finesse optical
cavity \cite{Mun02}.

\subsection{Lamb-Dicke Approximation}

The Lamb-Dicke approximation has been examined in some detail in
\cite{Parkins01}. Taking into account the spread of the phonon number
distribution associated with a general state, a condition for the
validity of this approximation can be derived as
\begin{equation}
\frac{1}{2}\eta_x^2\left(1+\bar{n}_x+a\sigma_{\bar{n}_x}\right) \ll 1 ,
\end{equation}
where $\bar{n}_x$ is the mean phonon number, $\sigma_{\bar{n}_x}^2$ is
the variance of the number state distribution, and
$a\sim 2-3$ (i.e., a few standard deviations from the mean).
If we assume that the number state distribution of each mode
is close to that of a thermal mode, then we can take
$\sigma_{\bar{n}_x}\simeq (\bar{n}_x^2+\bar{n}_x)^{1/2}\simeq \bar{n}_x+1/2$
for $\bar{n}_x>2$, and, with $a=3$, the condition becomes
\begin{equation}
\bar{n}_x \ll \frac{1}{2\eta_x^2} - \frac{5}{8} .
\end{equation}
For $\eta_x=0.1$ this reduces to $\bar{n}_x\ll 49$. As noted earlier,
at the time when minimum variances are generally achieved by our scheme
(e.g., with $\epsilon <1$ and $\lambda\geq 1$),
the mean phonon excitation numbers are of the order of 5--6, suggesting
that Lamb-Dicke parameters of the order of 0.1 or smaller are
sufficient.

Such values of the Lamb-Dicke parameter have been achieved with single
atoms in cavity QED settings \cite{Ye99,Gut01,Mun02}.
An ability to position the trap center very precisely at
any point along the cavity standing wave field
(e.g., at a node of the field, as required by the present scheme)
has also been demonstrated \cite{Gut01,Mun02}.

Before continuing, we briefly mention the further interesting possibility
of introducing a nontrivial time dependence to one or both of the coupling
laser fields.
One such example is illustrated in Fig.~8. Here, $\Omega_2=1$
(in dimensionless units) is fixed, while
$\Omega_1(t)=\Omega_2\sin (t/\tau )$ ($\tau =20,25,30$), i.e., the
effective parametric coupling strength is slowly increased from zero in
the first atom-cavity system.
If we consider the point for each case at which the variance is reduced
to 0.2 (i.e., a 90\% reduction below the ``vacuum'' level), then the
corresponding mean phonon number at this time is seen to be only of
the order of 2--3 for time-dependent $\Omega_1(t)$,
whereas for the case of constant $\Omega_1$ and $\Omega_2$ (also shown
in the figures) a variance of 0.2 is only attained with
a mean phonon number greater than 10.

We have not explored other time dependencies in detail, but it
would appear that there could be some advantage to doing so, in
particular, from the point of view of
minimizing the variance while maintaining mean phonon excitation numbers
which comfortably satisfy the Lamb-Dicke constraint.

\begin{figure}[h]
\begin{center}
\includegraphics[scale=0.42]{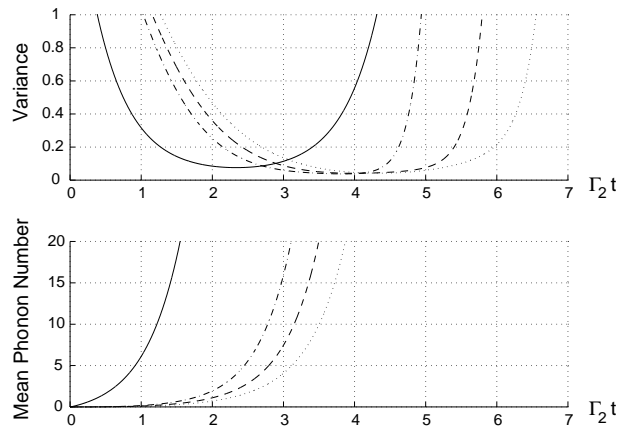}
\caption{Variance $\langle (X_1-X_2)^2\rangle$ and mean phonon number
$\langle b_1^\dagger b_1\rangle$ as a function of time from numerical
simulation of Eq.~(\ref{eq:mecas}) using
$\epsilon =1$, $\kappa_1=\kappa_2=10$,
$\Omega_2=1$ ($\Gamma_2=\Omega_2^2/\kappa_2=0.1$),
and $\Omega_1(t)=\Omega_2\sin (t/20)$ (dot-dashed line),
$\Omega_1(t)=\Omega_2\sin (t/25)$ (dashed line),
$\Omega_1(t)=\Omega_2\sin (t/30)$ (dotted line).
The case $\Omega_1=\Omega_2=1$ is shown as a solid line.}
\end{center}
\end{figure}

\subsection{Atomic Spontaneous Emission}

In practice, finite excitation of the atomic excited state will
introduce the effects of spontaneous emission to the dynamics.
Random momentum recoils associated with spontaneous emission
events will disturb the motional states in an uncontrollable
manner. However, if the rate at which such events occur is
small compared with the rate at which the entangled states of
interest are prepared by our scheme, then the effects of
spontaneous emission can essentially be neglected.

In \cite{Parkins01} the rate of spontaneous emission events
affecting the motional state is estimated for the kind of
configuration we are considering here. By comparing this rate
with the rates $\Gamma_i$, which characterize the preparation
of the entangled motional states, the general condition under
which the effects of spontaneous emission can be neglected is
derived to be
\begin{equation}
\frac{10g_0^2}{\kappa\gamma} \gg 1 ,
\end{equation}
where $\gamma$ is the linewidth (FWHM) of the internal atomic
excited state.
This is basically the condition of strong-coupling cavity QED,
as realized, for example, in the single-atom experiments of
\cite{Ye99,Hoo00,Pin00} (where values of $g_0^2/(\kappa\gamma )$
in the range 30--150 were achieved).

Finally, note that throughout this work we have neglected any
forms of motional state decoherence associated with the trap
itself. This seems reasonable, at least in the case of trapped
ions, where typical timescales for motional decoherence and
heating observed in recent experiments are of the order of
milliseconds or longer \cite{Roos99,Turchette00}.
With values of $g_0$ and $\kappa$ in the MHz range, we would
anticipate a timescale for preparation of the motional states
(i.e., $\Gamma_i^{-1}$) that is more than an order of magnitude
shorter than such decoherence times.

\section{Conclusions}

In conclusion, we have described a scheme for entangling motional
and light-field modes via an effective parametric (or two-mode
squeezing) interaction realized in a single-atom cavity QED
setting.
Through decay of the cavity field through one of its mirrors,
the cavity field entanglement is transferred to an external
propagating field which may be coupled to the motional mode of a
distant atom via a second cavity-QED-mediated interaction.
This enables the preparation of an EPR-type entangled state of
the position and momentum variables of separated atoms.
This has a distinct advantage over a related
scheme \cite{Parkins00a} for preparing such a state in that a
separate source of entangled light beams is {\it not} required.

We have given consideration to various practical issues associated
with the scheme and pointed to a variety of recent experiments in
single-atom cavity QED which collectively offer great encouragement
to our proposal.
Realization of the scheme would offer the exciting possibility
of implementing a variety of continuous variable quantum computation
and communication protocols, as well as opening the door to
further investigations of fundamental aspects of entangled quantum
systems, only now with distantly-separated {\it massive}
particles \cite{Parkins00a}.

\acknowledgments

AP was supported by a University of Auckland Summer Studentship.
ASP acknowledges support from the
Marsden Fund of the Royal Society of New Zealand
and thanks Lu-Ming Duan for discussions related to this work.

\end{document}